# A novel perspective on denoising using quantum localization with application to medical imaging

**Amirreza Hashemi**[1,*], **Sayantan Dutta**[2,3], **Bertrand Georgeot**[4], **Denis Kouamé**[5], and **Hamid Sabet**[1]

[1]Massachusetts General Hospital & Harvard Medical School, Department of Radiology, Boston, 02129, USA
[2]Weill Cornell Medicine, Department of Radiology, New York, 10065, USA
[3]Advanced Technology Group, GE HealthCare, Bangalore 560066, India
[4]Université de Toulouse, CNRS, UPS, LPT, Toulouse, 31042, France
[5]Université de Toulouse, CNRS, IRIT, Toulouse, 31042, France
[*]sahashemi@mgh.harvard.edu

**ABSTRACT**

Background noise in many fields such as medical imaging poses significant challenges for accurate diagnosis, prompting the development of denoising algorithms. Traditional methodologies, however, often struggle to address the complexities of noisy environments in high dimensional imaging systems. This paper introduces a novel quantum-inspired approach for image denoising, drawing upon principles of quantum and condensed matter physics. Our approach views medical images as amorphous structures akin to those found in condensed matter physics and we propose an algorithm that incorporates the concept of mode resolved localization directly into the denoising process. Notably, unlike previous studies that considered localization as a hindrance, our approach considers quantum localization as a fundamental component of image reconstruction which is used to differentiate between noisy and non-noisy modes based on diffusivity and localization measurements. This perspective eliminates the need for hyperparameter tuning, making the proposed method a standalone algorithm which can be implemented with minimal manual intervention and can perform automatic filtering of noise regardless of noise level. Through numerical validation, we showcase the effectiveness of our approach in addressing noise-related challenges in imaging and especially medical imaging, underscoring its relevance for possible quantum computing applications.

## Introduction

Imaging noise, such as speckle noise in ultrasound, Gaussian noise in magnetic resonance imaging (MRI), and sway of electronic noise, scatter radiation, random coincidences, and attenuation effects in nuclear imaging can obscure important anatomical structures and details, making it difficult for clinicians to accurately interpret images and make informed diagnoses. By effectively removing this noise, denoising algorithms enhance the clarity and fidelity of medical images, improving their diagnostic utility and enabling clinicians to identify subtle abnormalities or pathologies more accurately. In recent years, many computational techniques such as neural networks[1–4], regularization-based techniques[5–7], and statistical approach[8,9] have shown positive progress on addressing the challenges of image denoising and background noise reduction. However, despite the success of these methodologies, they exhibit deficiencies when confronted with the complexities inherent in noisy image environments. For instance, neural networks when dealing with learning intricate patterns and representations, often struggle with the detailed structures present in noisy images, especially in domains like medical imaging where data acquisition is constrained in imaging modalities due to design or target applications such as limited angle tomography[10–13] or Compton camera[14–16]. Furthermore, the reliance on large annotated datasets for training can be prohibitive, hindering the robustness and generalization capabilities of neural networks, particularly in the presence of significant noise levels. Additionally, neural networks suffer from overfitting, particularly when trained on small datasets or noisy images, leading to suboptimal denoising performance. Similarly, while regularization-based approaches provide a framework for controlling computational model and preventing overfitting, they may struggle to capture the diverse and complex structures present in noisy images due to non-linearity and inability to capture local extrema, or complex regularizations, thereby limiting their effectiveness in noise reduction tasks. These limitations underscore the need for alternative methodologies that can effectively address the complexities of denoising in high dimensional medical imaging while ensuring robustness and generalization across diverse imaging modalities and noise conditions.

In recent years, a few attempts have been made to apply quantum principles in image or signal processing, including early work[17] and proceeding efforts in image segmentation[18,19]. More recent developments[20–23] adopt a quantum inspired approach to imaging systems. These methods show promising start for the utilization of quantum physics into the denoising problems. One important aspect of those works was to process images as block-wise to preserve pixel correlation for efficient denoising, unlike previous methods that begin with a continuous mathematical representation and then discretize. Also, it was shown that the relevance of quantum localization phenomenon and quantum interference on noise level; however previous works did not use the quantum localization as a tool for denoising approach. Hence, like neural network and regularization approaches, this resulted that prior quantum-inspired methods rely on a set of hyperparameters regarding key parameters in both quantum mechanics theory and the filtering process, rendering them as quasi-automatic approaches requiring some manual intervention. Furthermore, in prior works[20–37], traditional filtering methods (such as hard or soft thresholding) were primarily used for denoising. However, these methods often led to the loss of localized wave-vectors crucial for finer imaging details (such as edges), resulting in a smoother but less detailed output. With the imminent rise of quantum computation, there is a growing need for standalone quantum approaches that effectively address these issues.

In this paper, we redefine the denoising approach emphasizing the central role of quantum localization in the denoising process. Here, we introduce a contrasting view of the medical image where we idealize a medical image as amorphous (disordered) structure akin to the condensed matter physics, and we use the amorphous model analysis to characterize the locality and propagating behaviors of the signals in decomposed imaging systems. In this view, we consider a noiseless image as a localized structure with the absence of diffusive behaviors, and therefore, in contrast, we consider the background noise as diffusive and non-local modal representatives. We will show that in various examples, similar characterization of the disordered regime is relevant in imaging space and is particularly useful for denoising. While the primary focus of this paper is on medical imaging, we also include examples of classical natural images to enable all readers to better evaluate its performance. Overall, the contribution of this paper lies in different points:
- We present a contrasting view where we draw a parallel between an image (e.g. medical image) and the theory of amorphous structures in condensed matter physics.
- Unlike previous studies that treated localization as a hindrance, we use localization properties to characterize the noise and differentiate it from the original signal.
- By interpreting vibrational modes in the amorphous regime, along with noise, as non-local modes, we introduce a thresholding and filtering approach that eliminates hyperparameter optimization, delivering a standalone quantum-inspired method.
- We introduce a novel and automated process to distinguish between local and non-local modes, utilizing the mode-resolved participation ratio and diffuson properties. By exploiting these quantum characteristics, we propose an efficient method to separate finer imaging details from noisy components, enabling a thresholding process grounded in the laws of physics rather than relying on traditional empirical thresholds.
- Lastly, the Planck constant which is also a hyperparameter of the model is estimated a priori from physical arguments without need of a process of optimization.
- It achieves a reduction in computational costs.

The paper proceeds with an organized examination of our proposition, beginning with an exposition on the basics of quantum mechanics. It then delves into modal analysis techniques applied to amorphous structures in condensed matter, offering insights into the characterization of vibrational modes. Next, the paper introduces quantum-inspired denoising methodologies, tailored specifically for image systems depicted as amorphous structures. Subsequently, it discusses the connection between the amorphous model and images affected by background noise, laying the groundwork for denoising in compressed sensing applications and filtering process. Through various examples, the paper demonstrates the effectiveness of these techniques in enhancing image quality and compressing the essential image components.

## Methods
### Basics of Quantum Mechanics

Quantum physics theory explores the behavior of particles at the quantum scales, challenging the classical understanding of the surrounding phenomena. This theory has been crucial for understanding the properties of complex systems such as solids, liquids, and gases. At the heart of quantum physics lies Schrödinger's equation, a fundamental equation that describes how the wave function of a quantum system evolves over time. In a non-relativistic single particle quantum system, a wave function $\psi(y)$ describes the probability of presence of a particle in a potential $V(y)$, where $y$ is the spatial position. This wave function is an element of a Hilbert space with bounded integrals and follows the stationary Schrödinger equation[24]:

$$H\psi(y) = E\psi(y) \quad (1)$$

where $H = (-\hbar/2m)\nabla_y^2 + V(y)$ is the Hamiltonian operator. Here $m$ and $E$ are the mass and the energy of the particle and $\hbar$ is the Planck constant that relates the energy of the particle to its frequency, $\nabla_y^2$ is spatial Laplacian derivatives at spatial positions $y$.

In condensed matter physics, the amorphous structure is often locally described by harmonic oscillators and the potential energy function takes the form of $V(y) = \frac{1}{2}m\omega^2 y^2$ where ω is the angular frequency of the oscillator. Substituting this potential into the Schrödinger equation yields the following:

$$\left((-\hbar/2m)\nabla_y^2 + 1/2\, m\omega^2 y^2\right)\psi(y) = E\psi(y) \quad (2).$$

This equation is known as the time-independent Schrödinger equation for the harmonic oscillator. It describes the energy levels and wave functions of a quantum harmonic oscillator system. The solutions to this equation give the quantized energy levels of the harmonic oscillator, which are equally spaced, and the corresponding wave functions represent the probability distributions of finding the particle at various positions along the oscillator. Thus, from the Schrödinger equation with the harmonic oscillator potential, we obtain a dynamical system of the quantum harmonic oscillator. The harmonic oscillator model can provide insights into the vibrational behavior of amorphous materials. In this context, the vibrational modes are not strictly phonons, but rather collective excitations involving the motion of atoms or molecules within the material. These excitations can still be approximated as harmonic oscillators, with each mode characterized by a specific frequency and associated energy.

In imaging examples, we assume that the pixels are the particles, and the potential is described by the intensity values of the pixels (i.e., for 2D image, $V = x$, where $x \in \mathbb{R}^{N \times N}$ represents the intensity value of image containing $N \times N$ pixels). Therefore, the Schrödinger equation yields the following form:

$$\left((-\hbar/2m)\nabla_y^2 + x(y)\right)\psi(y) = E\psi(y) \quad (3)$$

and similarly, this results in a dynamical system where the solution is the set of eigenvectors that serve as an adaptive basis in decomposed images or signals. As discussed in previous works, for a vectorized 2D image, when the conventional zero padding is used as boundary conditions for the Hamiltonian operator, the discretized Hamiltonian matrix, $H \in \mathbb{R}^{N^2 \times N^2}$, takes the simplified form of

$$H(i,j) = \begin{cases} x(i) + 4(\hbar^2/2m) & \text{for } i = j \\ -\hbar^2/2m & \text{for } i = j \pm 1 \\ -\hbar^2/2m & \text{for } i = j \pm N \\ 0 & \text{otherwise} \end{cases} \quad (4)$$

where $H(i,j)$ is the $(i,j)$-th component of the Hamiltonian operator. Unlike in quantum mechanics, the Planck constant in image processing is a parameter that should be tuned. Previous studies have optimized the value of the Planck constant manually to choose the optimal one in order to denoise an image. Here we propose a formula to estimate the optimal value based on quantum mechanics:

$$\hbar = \frac{E}{\nu} = \frac{2\left(\sqrt{\sum_{i=1}^{N^2}\left(x(i)/\max(x)\right)^2}\right)}{N} \quad (5)$$

where $E$ and $\nu$ are the energy and frequency of the decomposed image. Here the denominator represents the maximum frequency of an image denoted by a half of pixel number in one direction, $N/2$, and the numerator is denoted by the total normalized squared intensity of an image. Furthermore, we define $\hbar' = \alpha\hbar$ where $\alpha \in \mathbb{N}$ as an augmented value to study its influence on our problem.

**Amorphous Regime in Disordered Harmonic Solids**

In amorphous regime, materials exhibit a lack of long-range order in their atomic structure, setting them apart from crystalline solids. Within this context, the atomistic vibrational eigenmodes in amorphous materials can be classified into two main categories: propagating and non-propagating modes[25]. Propagating modes, characterized by longer wavelengths, are wavelike vibrational movements through a homogeneous medium and they are undiscerning of atomistic disorder. On the other hand, non-propagating modes are called diffusons and locons. Diffusons extend across the entire amorphous sample, representing vibrational eigenmodes that diffuse energy over the material without being confined to specific regions. In contrast, locons are spatially localized modes, where vibrational movements are trapped within specific regions of the material. In essence, the vibrational modes of atoms in amorphous materials can be broadly categorized into three types: propagons, which propagate and are non-local; diffusons, which do not propagate but are non-local; and locons, which are localized and non-propagating[25–27]. Figure 1 demonstrates the behaviors of these three vibrational modes in amorphous structure.

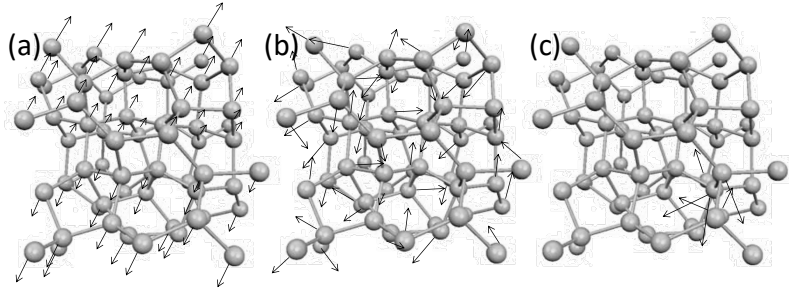

**Figure 1.** Demonstration of vibrational modes in amorphous structures: a) propagon, b) diffuson c) locon.

The distinction criteria of the vibrational modes in amorphous structures correspond directly to the structural properties of the disorder. For distinction between propagons and diffusons, a well-known Ioffe-Regel criterion[28] comes into play. According to this criterion, as the mean free path of a particle, such as a phonon, approaches the magnitude of the interatomic spacing or the size of the material's structural units, a transition occurs in the vibrational modes from propagons, indicative of propagating states, to diffusons, representing diffusive or non-propagating states. For diffusons-locons separation, Anderson localization[29] provides a useful picture which refers to the phenomenon where wavefunctions become localized in a disordered medium, preventing the propagation of waves over long distances. In the case of vibrational modes in amorphous materials, Anderson localization can lead to the trapping of vibrational energy within specific regions due to the disorder in the atomic arrangement. The distinguishing criterion for separating diffusons from locons is commonly referred to as the mobility edge[25].

In terms of computational tools, the participation ratio serves as one of the indicators of localized modes and Anderson localization phenomena in amorphous materials. A low participation ratio means that an eigenstate is highly localized, indicating that its amplitude is concentrated within a small number of elements. Conversely, a high participation ratio implies that the eigenstate is more spread out or delocalized, with its amplitude distributed across numerous elements. For vibrational modes, the participation ratio is given as follows:

$$\text{Pr}_n = \frac{\left(\sum_b \vec{e}_{b,n}^2\right)^2}{N \sum_b \vec{e}_{b,n}^4} \qquad (6)$$

where $\vec{e}_{b,n}$ is the eigenvector for mode $n$ at atom $b$ and $N$ is the total number of atoms.

To demonstrate the relevance of the participation ratio and criteria for distinguishing vibrational modes in amorphous structures, we computed participation ratio for a common amorphous Silicon (a-Si) structure that inhibits the short-range order (SRO), which means a length scale smaller than 5 Å, while lacking long-range order[30]. We employed a previously developed continuous random network (CRN)[31] as an illustrative example. The atomistic structure generated from the CRN utilizes a random-based atomic arrangement method with a bond-swapping algorithm. The CRN framework constructs the structure with SRO and preserves disorder beyond the second neighbor lengths, resulting in the elimination of defects and voids. Specifically, the CRN structure of a-Si exhibits a defect and void concentration of less than 1-3%. The simulated system contains 4096 atoms and the Tersoff potential[32] is used in GULP package[33] to obtain eigenvectors by solving the dynamical system. Figure 2 shows the participation ratio for example a-Si structure.

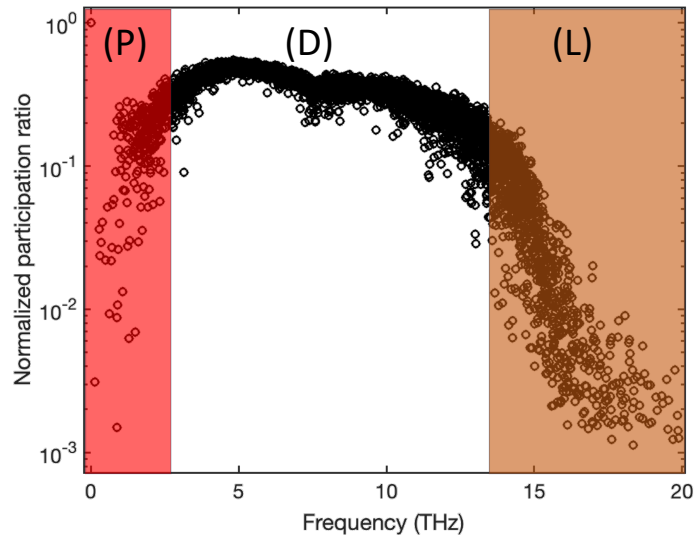

**Figure 2.** Mode resolved participation ratio in a-Si; regions P, D, and L correspond to propagons, diffusons, and locons.

The two shaded regions indicate the separation of propagons at low frequencies from diffusons and diffusons from locons at higher frequencies. While diffusons dominate the frequency range the propagons and locons count for small fraction of the total modes. We should note that the separation criteria are subjective and usually multiple factors are considered to obtain a clear-cut separation. However, the calculation for participation ratio is straightforward with relatively low computational cost and this behavior is rather consistent between all amorphous structures.

## Results and discussions

### Amorphous Regime and Localization in Image Processing

Images are broadly considered as spatially structured data with localized patterns. This prompted us to explore whether mode resolved localization characterization tools are applicable for addressing medical imaging problems and other scenarios where spatial localization within the image is a key aspect. Here we investigate the presence of localization in image with noise and noiseless structures using the concept of mode resolved participation ratio, inspired with the analogy with amorphous solids. Similar to condensed matter analysis of vibrational modes, we calculate the participation ratio of the sample image as following:

$$\Pr_n = \frac{\left(\sum_i \vec{e}_{i,n}^2\right)^2}{N \sum_i \vec{e}_{i,n}^4} \qquad (7)$$

where $\vec{e}_{i,n}$ is the eigenvector for mode $n$ and pixel number $i$ and $N$ is total number of pixels.

The top row of Figure 3 shows the benchmark synthetic image of the size 64×64 pixels (which results in the same total number of modes as a-Si in Figure 2.) and noisy images with Poisson noise at signal to noise ratio (SNR) of 2, 5, and 15. The eigenvectors and eigenvalues of the synthetic image and noisy ones were calculated from dynamical system, equations 3 and 4, at estimated Planck constant of 0.7864 and subsequently the normalized participation ratio, the normalization is defined as the participation ratio over the total number of modes (pixels). The bottom plots in Figure 3 show the participation ratios for the corresponding synthetic image and noisy ones.

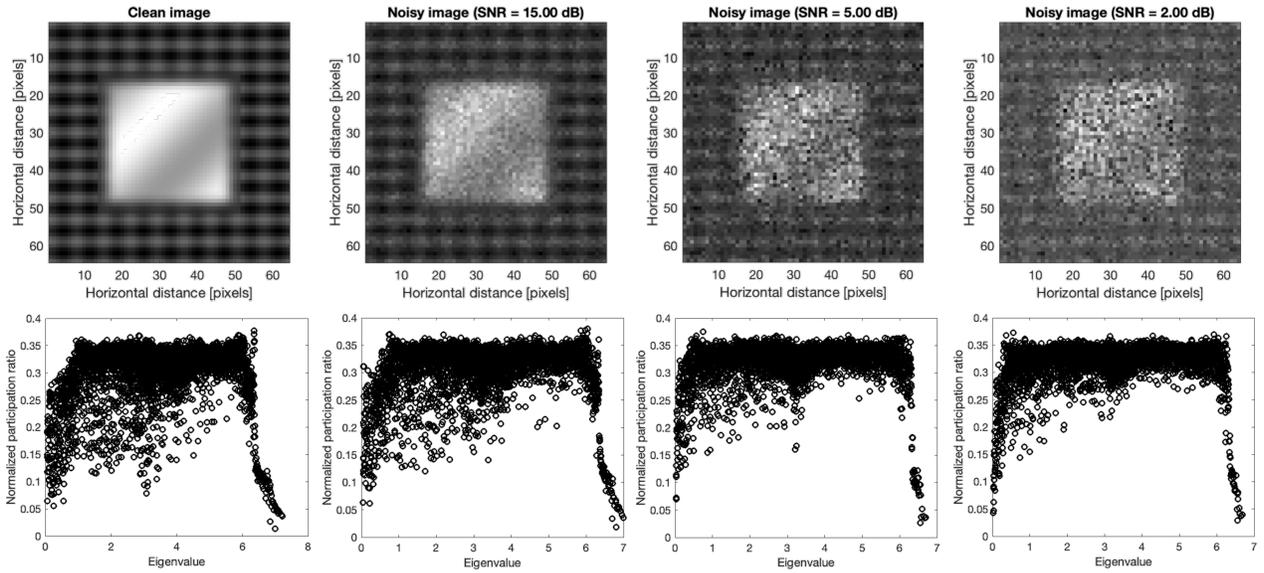

**Figure. 3.** Benchmark synthetic image, and noisy images with Poisson noise and SNR of 2, 5, and 15 are shown in the top. The bottom shows the corresponding the normalized participation ratio [normalized by the total number of pixels] plots of the clean and noisy images.

Interestingly the pattern of localization and participation ratio resemble the similar behavior as vibrational modes in amorphous regime. The comparison between participation ratio results shows that as SNR increases, the degree of the localization is diminished, and the mode resolved participation ratio uniformly increases particularly for the mid-range eigenvalues. In parallel, the identification of localized modes becomes more and more clear for low and high eigenvalues as SNR increases. This implies that the increase in noise level increases the participation of the non-local modes and ultimately results in a more clear-cut separation of the low and high localized modes. For large noise, e.g. SNR=2, there is a clear localization of the low and high eigenvalues, and the mid values dominate the eigenvalue domain. To further demonstrate the similarities between the vibrational modes in amorphous regime and mode resolved image, we show the plots of selected eigenvectors through three cross-sections [corresponding to the first to third rows of the image] for low, mid, and high eigenvalues in Figure 4.

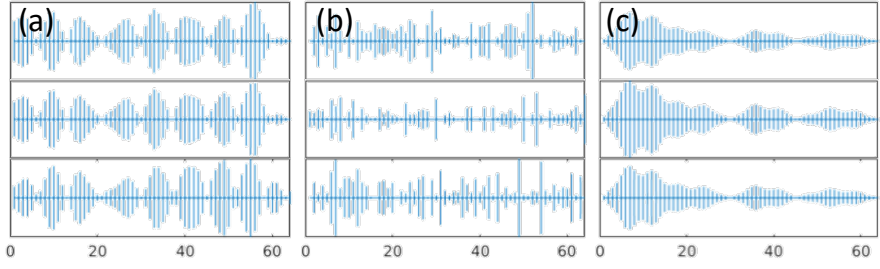

**Figure 4.** Sample eigenvectors of the decomposed image signal for SNR=2 at (a) low, (b) mid, and (c) high eigenvalues. Each row shows three cross-sections of an eigenvector.

Three cases show the propagating and non-local behavior for the low range modes, scattered and non-localized behavior for mid-range modes and, localized with non-propagating nature for high-range modes. Therefore, with comparison with Figure 1, the eigenvectors of imaging system are comparable to amorphous structures in condensed matter physics. We should note that although the same characteristics are apparent between the mode resolved image regime and vibrational modes in amorphous regime, the similar naming of propagon, diffuson and locons may not be appropriate in studying imaging systems. Also, we emphasize that the quantum localization premise in imaging systems relies on the spatial localization of the image's structural components. Hence, this characteristic can be effectively utilized for image processing. In the next section, we will discuss the connection between localization and background noise, and we propose our approach of utilizing localization to remove noise and compress the image reconstruction.

**Automated Denoising and Compressed Sensing via Quantum Localization Filtering**

Besides the environmental conditions and the limitations on the measurement systems, background noise often refers to unwanted signals or interference that diffuses through the image due to the scattering of incident radiation or waves by particles or structures within the imaged area. Hence, we consider background noise as a non-localized mode that is participant in the characteristics of many pixels, and they represent the scattered behavior similar to the diffusons in vibrational modes of amorphous structures. To examine this proposition, we use a modified version of the quantum inspired approach[20] for denoising process that was constructed by smoothing the input signal, computing the eigenmodes, manually thresholding them, and then back-projecting the thresholder eigenmodes. In this approach which we call quantum localization approach hereafter, we introduce a criterion to remove/filter the non-localized modes with high participation ratio based on the histogram distribution plot of the normalized participation ratio. Given that the majority of modes are non-local modes with high participation ratio, we fit a Lorentzian function on the spectral distribution to select these non-local modes. The Lorentzian function, $\Phi$, as a function of participation ratio is defined as

$$\Phi(\lambda) = \frac{1}{\pi}\left[\frac{\Gamma}{[\lambda-\lambda_0]^2+\Gamma^2}\right] \qquad (8)$$

where the $\lambda_0$ is the peak participation ratio value, $\Gamma$ is the half width at half maximum (HWHM). Figure 5 shows the distribution of the participation ratio for the noisy synthetic image of SNR=2 and fitted Lorentzian function on the peak distribution in the outset. The corresponding region below the Lorentzian function fit gives a good approximation of the high participation and non-local modes. Here we consider the lower bound of the Lorentzian fit and the eigenvalues corresponding to the low and high values as the threshold to filter the signals for both low and high range modes. The inset of the figure 5 shows the selected region that is used in the localized approach.

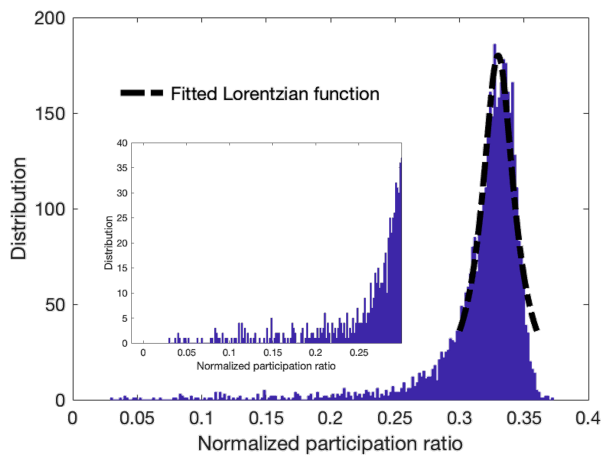

**Fig. 5.** Outset plot shows the distribution of the normalized participation ratio of the synthetic image with the fitted Lorentzian function at the peak of the distribution. The inset plot shows the zoom-in view of the selected regions that was used to extract the localized modes at low and high eigenvalue ranges.

This results in fixing an effective value for the thresholding in the adaptive basis, which is in this way determined from the noisy image directly. The summary of presented modifications is summarized in Algorithm 1.

| Algorithm 1: denoising algorithm using the quantum localization approach |
|---|
| **Input**: Image, $x$ |
| 1. Compute the Planck constant using equation (5) and Hamiltonian matrices, $H$, using equation (4) |
| 2. Calculate the eigenvectors of equation (3) and singular value decomposition (SVD) |
| 3. Calculate the participation ratio using equation (7) |
| 4. Fit the Lorentzian fit using equation (8) to the distribution of normalized participation ratio and threshold based on the image modes with non-local and non-propagating modes |
| 5. Recover the denoised image using the filtered image |
| **Output**: Denoised image, $x'$ |

Figure 6 shows the denoised images using the full set of eigenvectors based on the previous work[20] and filtered denoised images are the results of the quantum localization approach using only low and high range modes which are shaded in the corresponding participation ratio of noisy image, these results are shown for three SNR's of 2, 5, 10, and 15. Additionally, histograms of the normalized participation ratio are provided for further clarification on its behavior.

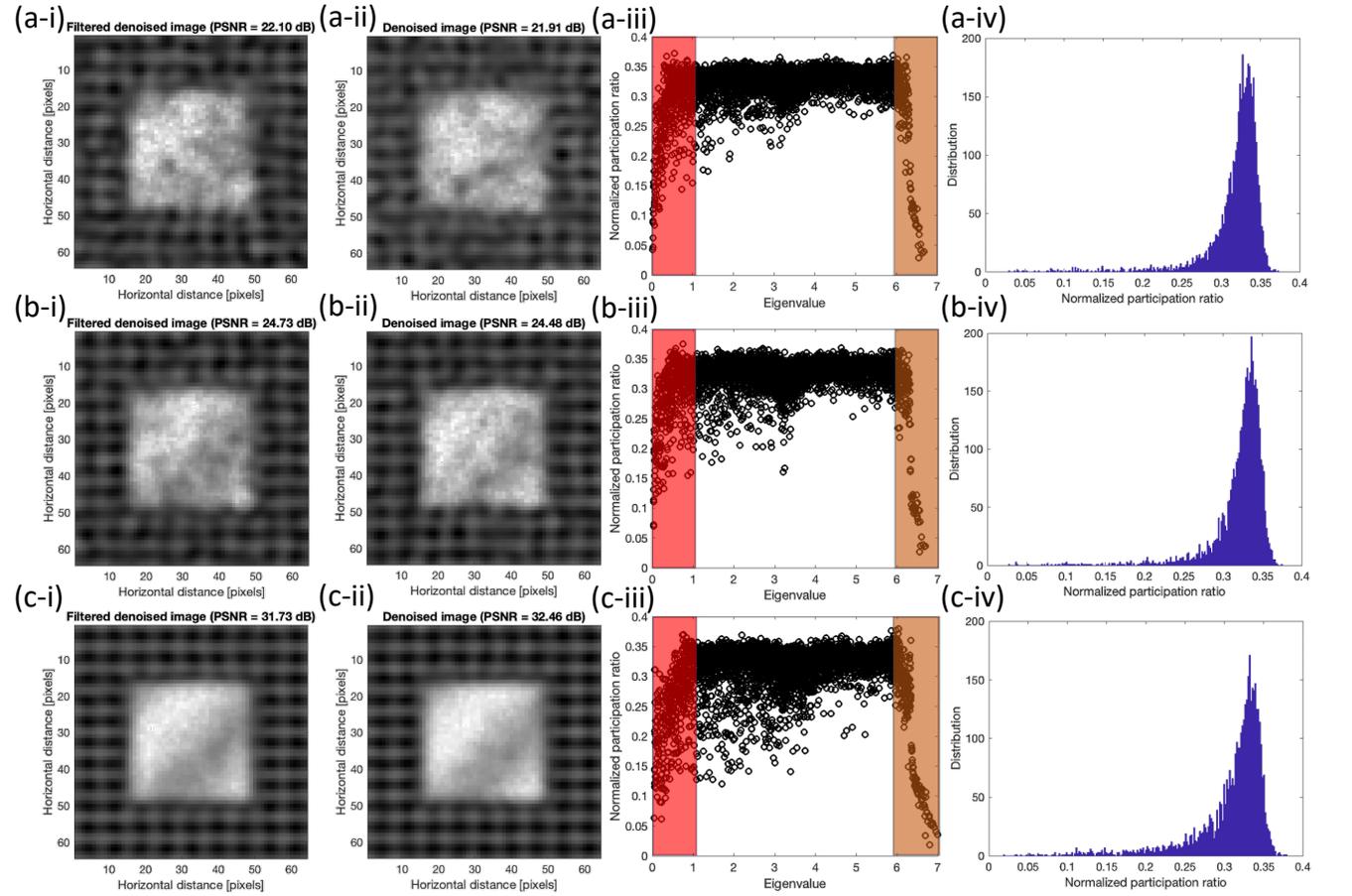

**Figure 6.** Comparison between the denoised results of the (i) quantum localization and (ii) original approaches at SNR's of (a) 2 (b) 5 (c) 15. Corresponding normalized participation ratio plots with the shaded regions which indicate the filtered regions in localized approach and corresponding histogram of the normalized participation ratio are shown in (iii) and (iv) respectively.

In all cases, we observe the comparable results between the quantum localization approach, which include only shaded region, versus the approach with all mode inclusion. To clarify the differences Table I shows the comparison of SSIM and

PSNR between two methods.

| SNR | SSIM | | PSNR (dB) | |
|---|---|---|---|---|
| | All modes | Selected modes | All modes | Selected modes |
| 2 | 0.5 | 0.5 | 22.1 | 21.91 |
| 5 | 0.65 | 0.64 | 24.73 | 24.48 |
| 15 | 0.89 | 0.9 | 31.73 | 32.46 |

**Table 1.** Comparison SSIM and PSNR for the synthetic Image

The values of SSIM's are slightly smaller for lower SNR values while slightly larger for the higher SNR values. Overall, the comparison between two approaches, filtered and non-filtered, shows that the contribution of the mid-range eigenvectors is minimal in the reconstruction process and the majority of those modes are related to the added noise. Also, we observe that as the noise level increases, the efficacy of the filtering becomes more effective.

**The Role of Planck Constant on Localization in Imaging System**

The Planck constant is not a well-established concept for imaging systems. To understand the influence of the Planck constant on the localization, we calculated the participation ratio based on the augmented Planck constants, $\hbar' = \alpha\hbar$, as assumed for $\alpha = 0.25, 0.5, 1, 2,$ and $4$ [where for $\alpha = 4$ the $\hbar'$ is near to the previously used Planck constant]. Figure 7 shows the corresponding results for the synthetic image with SNR=2, for the augmented Planck constants.

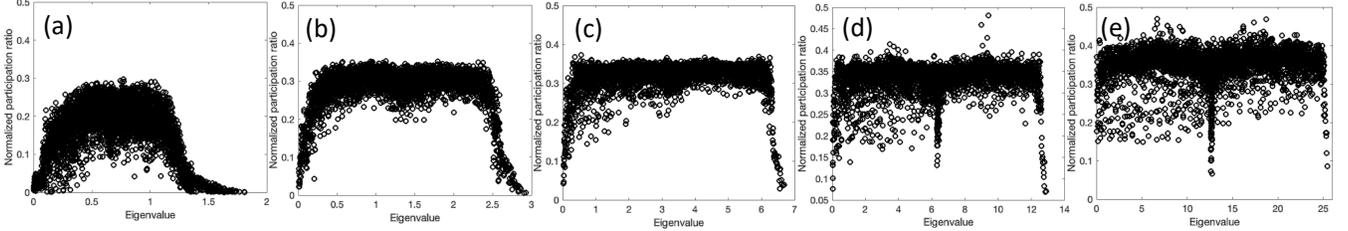

**Figure 7.** Normalized participation ratio plots of the synthetic image with SNR=2 with augmented Planck constant $\hbar'$ for (a) $\alpha = 0.25$ (b) $\alpha = 0.5$, (c) $\alpha = 1$, (d) $\alpha = 2$, and (e) $\alpha = 4$.

The results show that the localization or the distinction between different regimes is largely diminished as Planck constant increases. In physics sense, a large Planck constant implies that the Planck length can become larger than the corresponding localization length. For smaller Planck constants, we observe mid-range frequency modes collapsing to low participation ratio values, indicating that the choice of the Planck constant itself contributes to localization rather than the image intensity values used in constructing the Hamiltonian matrices. Our results show that the choice of Planck constant plays an important role in the identification of the different modes.

**Results for the complex image**

Here, we present the results for a complex image (Lena image) with a size of 256×256 pixels. We compare the outcomes obtained through the original method, where all modes are utilized for the denoising process, with those achieved through the quantum localization approach, which involves selecting specific modes for denoising and reconstruction, as shown in figure 8. The added noise follows a Poisson distribution with an SNR of 15.

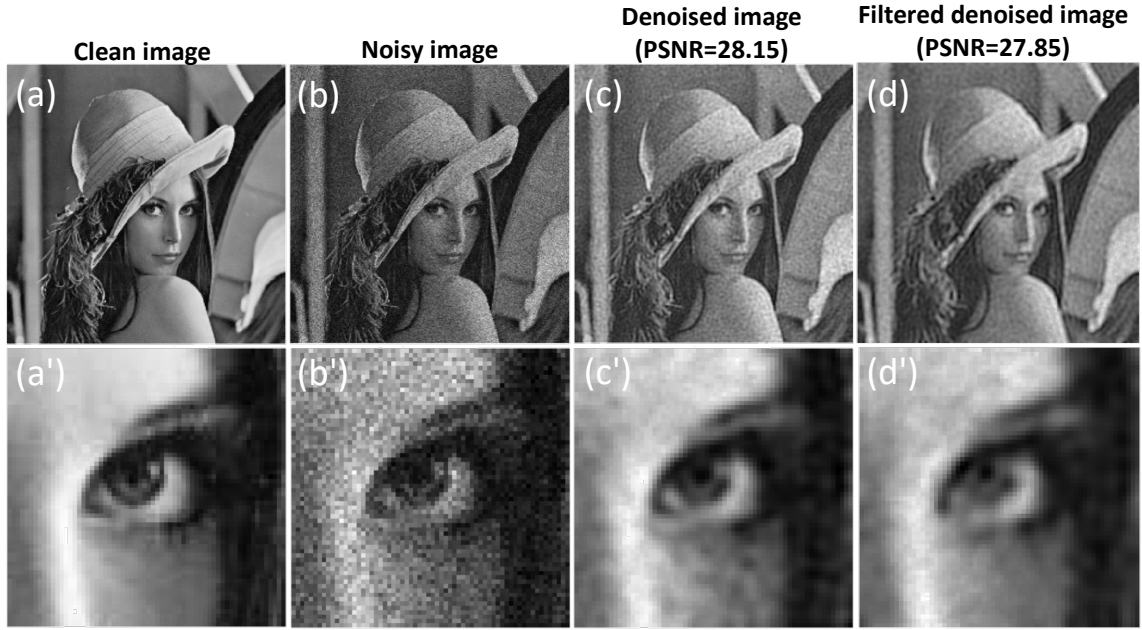

**Figure 8.** Comparative analysis of denoising methods: quantum inspired approach with all modes vs. quantum localization approach with selected high localization modes. The figure illustrates (a) the Lena image and its zoom-in counterpart (a'), (b) the noisy image and its zoom-in counterpart (b'), (c) the denoised image using all modes and its zoom-in counterpart (c'), and (d) the denoised image using the quantum localization approach and its zoom-in counterpart (d').

The results illustrate that the denoising process yields comparable results for the quantum localization version when compared to considering all modes. Specifically, the denoising results exhibit a slight decrease in contrast, accompanied by a slightly smoother appearance compared to the original denoising approach. The SSIM values for the original and quantum localization approaches are 0.76 and 0.75, respectively, while the PSNR values for the previous and quantum localization approaches are 28.15 dB and 27.85 dB, respectively.

Additionally, Figure 9 shows the normalized participation ratio for the noisy Lena image presented in Figure 8-(b). The shaded regions represent the corresponding low and high eigenvalue ranges based on the Lorentzian fit of the participation ratio distribution following the quantum localization approach. We observe a nearly symmetrical participation ratio between low and high eigenvalues, with greater localization and distinct separation into the high participation ratio (mid-eigenvalue) region. These results suggest that for complex images, localization is notably more pronounced compared to simple structured images.

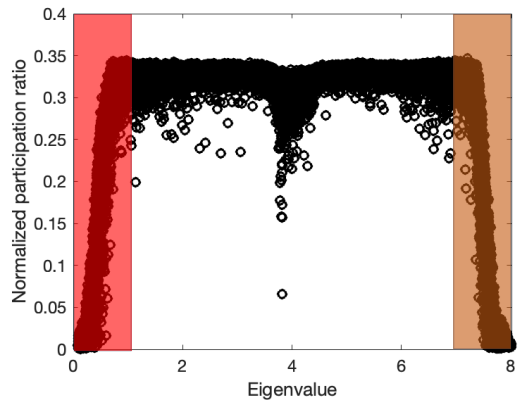

**Figure 9.** Normalized participation ratio plots of noisy Lena images. Shaded regions approximately show the corresponding low and high eigenvalue parts that are truncated in the quantum localization approach.

**Results for the medical image**

Lastly, we compare the results obtained between original and quantum localization methods for a computed tomography (CT) image with a size of 260×260 pixels in Figure 10.

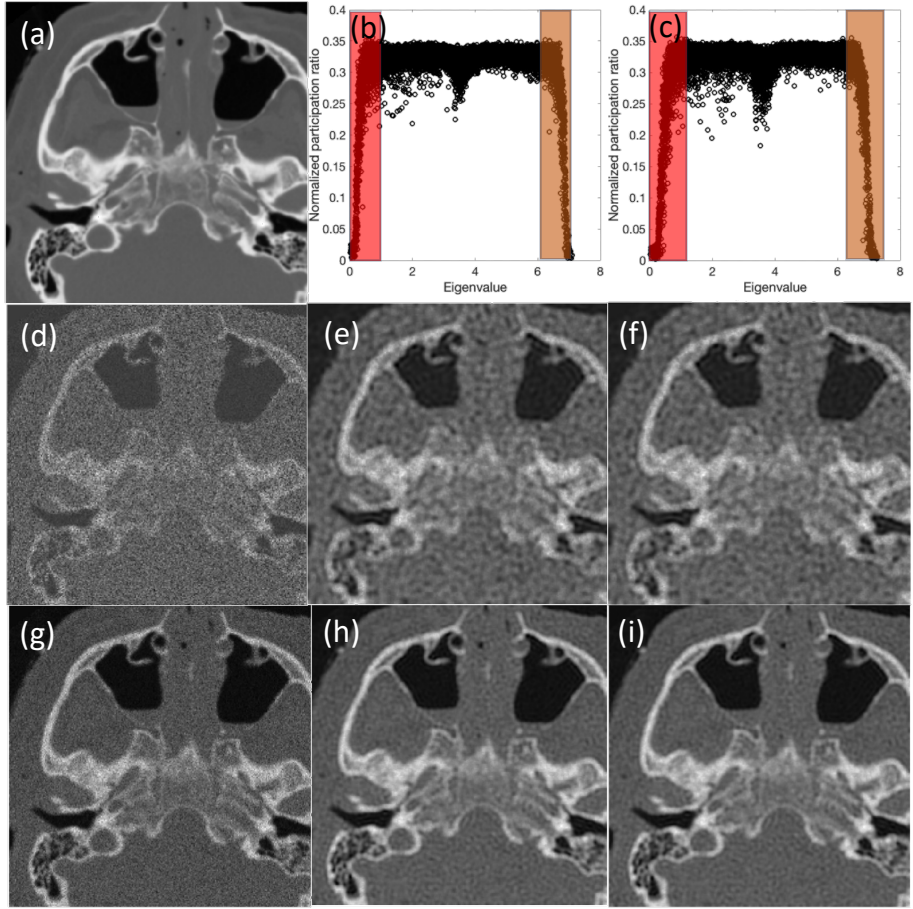

**Figure 10.** Comparative analysis of denoising methods: quantum inspired approach with all modes vs. quantum localization approach with selected high localization modes for CT Images. The figure depicts (a) the original CT image, (b) normalized participation ratio vs. eigenvalues for a noisy CT image with SNR=5, and (c) normalized participation ratio vs. eigenvalues for a noisy CT image with SNR=15; shaded regions show the low and high eigenvalues that are selected in the quantum localization approach. For SNR=5, the figure displays (d) the noisy CT image, (e) the denoised image using all modes, and (f) the denoised image using the quantum localization approach. For SNR=15, it shows (g) the noisy CT image, (h) the denoised image using all modes, and (i) the denoised image using the quantum localization approach.

The results show the success of the quantum localization approach to capture the comparative results in all cases while about 70% of all mid-range eigenvalues are dismissed. To compare two approaches in terms of quality metrics, Table 2 summarizes the SSIM and PSNR values; those values consistently show the excellent performance for the quantum localization approach and point to the success of the criteria to choose the effective modes based on the quantum localization for the medical image example.

|     | SSIM      |                | PSNR (dB) |                |
| --- | --------- | -------------- | --------- | -------------- |
| SNR | All modes | Selected modes | All modes | Selected modes |
| 2   | 0.51      | 0.51           | 22.73     | 22.73          |
| 5   | 0.57      | 0.57           | 24.62     | 24.60          |
| 10  | 0.69      | 0.69           | 27.81     | 27.81          |
| 15  | 0.79      | 0.79           | 30.84     | 30.85          |

**Table 2.** Comparison SSIM and PSNR for the CT Image

Additionally, we tabulated the simulation times of both approaches for the CT image in Table 3. Each simulation was conducted using in-house MATLAB code on a single node of Bridges2 (with the regular memory 256 GB) at the Pittsburgh Supercomputing Center, and a cluster with 2 AMD EPYC 7742 CPUs, with each node boasting 64 cores and 128 threads. The results demonstrate a consistent computational time savings of over 30% across all SNR cases. We should note that the computational performance of the quantum localization approach also influences the compression of the decomposed image and enhances the management of memory allocation. Although we have demonstrated the performance improvement using the quantum localization approach even on classical computing units such as CPU supercomputers, but the full extent of the computational performance could be prominent when applied on quantum computing devices.

| SNR | Computational time selected modes (hrs) | Computational time all modes (hrs) |
|---|---|---|
| 2 | 30.78 | 45.75 |
| 5 | 38.08 | 55.69 |
| 10 | 55.21 | 84.9 |
| 15 | 75.46 | 109.18 |

**Table 3.** Computational times for the CT Image

## Remarks

In this paper, we compared an imaging system to an amorphous structure in condensed matter physics and used the idea of localization directly to deal with the noisy image. We proposed an algorithm as a revision of previously developed quantum inspired approach for denoising that utilizes the concept of the localization for filtering process. The quantum localization approach has comparable results to the original method while it compresses the decomposed imaging modes by over 70 %. This method also demonstrates superior denoising performance compared to traditional approaches such as total variation (TV), wavelet, and deep learning (DL) methods. For example, in the case of a synthetic image with SNR = 15, the PSNR values[20] achieved by the TV, wavelet, and DL approaches are 26.23 dB, 25.68 dB, and 27.22 dB, respectively—all of which are lower than the PSNR value achieved by the quantum localization approach presented in this study.

One important advantage of the quantum localization approach is the independence to the set of hyperparameters which were involved in the filtering process. The Lorentzian function proves to be a robust fit for describing the spectral lines of distribution plots and is effective to determine the separation lines in participation ratio vs eigenvalue plots.

In regard to Planck constant definition, the goal was to select the smallest value to capture all quantized image elements. Based on our observations, our definition provides a sufficiently small value where the localized modes are identifiable and the denoising process performs with a good quality. We should note that even smaller value Planck constant results in high localization even for mid-range eigenvalue modes and considerably lower quality denoised images. Despite this trade-off, we maintain confidence in the suitability of our chosen Planck constant definition for our imaging system.

We note that in previous work[20] a smoothing process on the image was used in order to compute the adaptive basis. This process was necessary when Anderson localization due to high noise levels leads to all wave functions being localized on a small part of the system and will be difficult to couple with the present approach. This smoothing process was especially needed for one-dimensional signals, even with relatively moderate levels of noise. For two-dimensional systems like images, this regime is present but only at very high noise values, and the present procedure can be safely used up to quite high values of noise.

The quantum localization approach is a standalone algorithm with minimal manual handling of the variables. As mentioned in the recent literature[34,35], achieving quantum level performance using conventional classical algorithms such as machine learning may not be tenable on future quantum computers. This exemplifies the importance of this work where the quantum inspired approach solely relies on quantum physics basics with minimal appeal to conventional classical algorithms such as regularization and machine learning. Additionally, the concept of localization is not limited to the participation ratio characterization, as discussed in previous works[25,27], there are number of other tools to characterize the amorphous model that might be relevant and applicable to the imaging systems. We purposefully limited the current algorithm to the use of participation ratio due to simple required modifications, and low computational cost. It is worth noting that the current modifications and representation of the imaging system as an amorphous model are highly adaptable and can be integrated with previous advancements in quantum-inspired approaches[36–39,49]. This integration has the potential to yield more efficient algorithms for medical imaging[50–51], both qualitatively and computationally.

## Conclusions

In conclusion, this paper presents a novel quantum-inspired approach for denoising images, leveraging the concept of quantum localization within an amorphous model framework. Building upon previous work, the proposed algorithm demonstrates comparable performance to existing methods while significantly reducing the computational complexity and eliminating the need for manual intervention in parameter tuning. Furthermore, the adoption of quantum principles aligns with the growing interest in quantum computing and offers promising avenues for addressing noise-related challenges across diverse imaging modalities both on system level and component level. The former refers to reconstructed image resolution in nuclear medicine modalities especially in count starved geometries such as limited angle tomography using time of flight Positron Emission Tomography[10,11,40,41], organ-specific Single Photon Emission Computed Tomography[14,42,43]. On detector level implementation, quantum-inspired denoising can play a significant role in nuclear imaging detectors during event positioning estimation[44], and detector projection processing for novel unconventional detector designs[45–48]. By embracing quantum principles without heavy reliance on additional classical algorithms, our standalone approach presents a promising path for using quantum-inspired tools in image processing and could lead to new quantum computing algorithms in the line of recent proposals[38].

## Acknowledgments

We acknowledge support from NSF grant: med240002p, enabling computational simulations through the ACCESS program clusters. This work is partially supported by the NIH grant R01HL145160. This work was also supported by CIMI Excellence Laboratory under ANR Grant ANR-11-LABX-0040 within the French State Programme "Investissements d'Avenir." This work was conducted while S. Dutta was affiliated with the Department of Radiology, Weill Cornell Medicine, New York, NY 10065, USA.

## Author contributions statement

Conceptualization, A.H.; methodology, A.H., B.G., D.K., H.S.; formal analysis, A.H., S.D.; data curation, A.H., S.D; Supervision, B.G., D.K., H.S.; writing—original draft preparation, A.H.; writing—review and editing, A.H., S.D., B.G., D.K., H.S. All authors read and approved the final manuscript.

## Competing interests

The authors declare no competing interests.

## Additional information

Correspondence and requests for materials should be addressed to A.H.